# THE REMOVAL OF NUMERICAL DRIFT FROM SCIENTIFIC MODELS

John Collins[1,2], Brian Farrimond[1,2], David Flower[1], Mark Anderson[2] and David Gill[3]

[1]SimCon Ltd, Dartmouth, United Kingdom
john.collins@simconglobal.com

[2]Department of Computing, Edge Hill University, Ormskirk, United Kingdom
mark.anderson@edgehill.ac.uk

[3]National Center for Atmospheric Research, Boulder, Colorado, USA

gill@ucar.gov

*Abstract*

*Computer programs often behave differently under different compilers or in different computing environments. Relative debugging is a collection of techniques by which these differences are analysed. Differences may arise because of different interpretations of errors in the code, because of bugs in the compilers or because of numerical drift, and all of these were observed in the present study. Numerical drift arises when small and acceptable differences in values computed by different systems are integrated, so that the results drift apart. This is well understood and need not degrade the validity of the program results. Coding errors and compiler bugs may degrade the results and should be removed. This paper describes a technique for the comparison of two program runs which removes numerical drift and therefore exposes coding and compiler errors. The procedure is highly automated and requires very little intervention by the user. The technique is applied to the Weather Research and Forecasting model, the most widely used weather and climate modelling code.*

## 1. INTRODUCTION

The technique of debugging by identifying differences in execution between versions of a program was introduced by Abramson and Sosic in 1994 and given the name relative debugging [1][2][3]. A relative debugging system known as Guard was developed that enable users to run debuggers on several versions of the program simultaneously [4]. Commands are provided that enable users to identify data structures within the program to have breakpoints set at which the contents of the data structures are compared, with optional choice of tolerance of insignificant differences. The user can analyse the differences using text, bitmaps or more powerful visualisation tools. The system is able to compare versions written in different languages - the user identifies the data structures to be compared through variable name and location within the source code files.

The importance of this techniques lies not with the development of new software, but with the on-going evolution and maintenance of existing software systems [2]. Whilst debugging plays an important role in the implementation of new systems, the ability to make comparisons between versions of software, and specifically the data being processed within the code, provides a powerful mechanism for ensuring that the software is delivering consistent and correct results throughout its existence. This is achieved through the comparison of values between a reference program and a development program (Figure 1).

Figure 1: Relative Debugging [2]

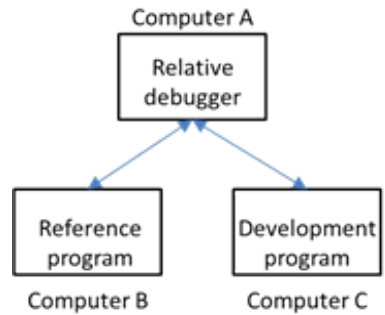

Abramson et al [2] describe the successful application of a relative debugging technique for a program running in a sequential and a parallel environment. Their technique requires the user to specify specific data structures to be compared during the different program runs. A difficulty of the technique is that numerical drift will eventually cause any assertion that the data structures are similar to fail. Also, there may be appreciable labour in choosing and specifying the appropriate data structures. Both issues are addressed in the technique described within this paper.

## 1.1. Numerical Drift

Numerical drift between runs of the same program in different computing environments occurs because of the integration of small differences in the results of floating point computations. These differences are to be expected. There are, at least, three reasons why they should occur:

i. different systems may use different floating point representations for numbers with different precision. The results of floating point computations are therefore often slightly different. Differences from this cause are now relatively uncommon because most computers use the same format of IEEE floating point numbers;

ii. different compilers may make different choices in the order of evaluation of terms in floating point expressions. The order of evaluation is usually strictly defined by, for example, the Fortran standard [5]. However, optimising compilers may sometimes violate the standard in the interests of efficiency;

iii. different compilers may make different choices of variables or quantities to be held in the processor registers. The processor registers of a typical PC chip are 10 bytes long. Values written to memory are 8 or 4 bytes long. The additional precision of values held in registers may be integrated to cause drift.

The objective of this study is to remove numerical drift. Any remaining differences between environments are likely to be the result of coding or compiler errors.

## 2. RELATED WORK

The Guard system was developed as a means of implementing automated relative debugging [2]. The project was driven by the fluid nature of software development for scientific modelling in which a community contributes towards the on-going development of the model. The community would generally consist of academics who have a different programming philosophy from those who program on engineering models [6], and thereby require a mechanism to ensure that the results generated by incremental changes to the model implementation do not contain significant differences.

The implemented system was successful in the hands of those users familiar with the code and likely causes of error. However, it suffered from a number of drawbacks which the system described in this paper aims to overcome. It has been argued that the automation in the Guard system is somewhat limited [7], in that Guard is deployed to perform comparisons. Further extensions have been proposed which extend the functionality to backtrack and re-execute code [8][9].

In looking for possible bugs, the user declares assertions to specify which data structures to monitor and where in the code they are to be compared (Abramson et al. 1995). The point in the code where a comparison is made that result in significant differences is unlikely to be the point where the differences are being generated. If differences are indeed detected then further work is required to pin down the cause of the differences, for example by introducing further monitoring points. The user's experience and intuition are required to make the detection of the location of the generation of differences efficient. Visualisation tools are used to help the user interpret and identify the cause of differences in these circumstances.

The differences may be spurious; for example, the result of numerical drift between the executions of the two versions. Valid discounting of the effects of numerical drift depends on the experience of the user. In contrast the system described in the current paper automatically removes the effect of numerical drift.

Later improvements to the Guard system (given the name DUCT) have focussed on attempting to automate the creation of assertions aimed at identifying the cause of differences[10][11]. The variables that are used to generate the values in the variable under investigation are recognised as the source of the differences. They are automatically identified and chains of preceding variable assignments that contribute to the values they hold when being used to modify the target variable are created. From these chains, further assertions are generated. Although this technique moves the user to the source of the generation of differences it still depends on the user's intuition regarding potential data structures where differences might occur. A technique that systematically examines every variable's assignment is capable of detecting differences caused by errors not expected by the user. This would be a full automation of the procedure rather than the hybrid presented by DUCT. The system described in this paper provides such a technique by identifying automatically the locations in the code where all significant differences are generated.

The Guard system is able to compare programs written in different languages because the user explicitly identifies the data structures in each to be compared and where in their source code files to compare them. The system described in this paper works only for programs written in Fortran since it depends upon a semantic analysis and automatic code modifications provided by the tool kit, WinFPT .

An alternative relative debugging project, HPCVL [7], has considered the limitations in Guard relating to the identification of equivalent variables in the reference and development programs. As identified, the variable names may differ between the two versions of the code which would make the comparison difficult, if not impossible. In order to overcome this, a template which adopted a labelling mechanism was implemented. This required the source code to be manually amended to include calls to library functions which could be included or removed by pre-processing the code. Whilst this provides a powerful means by which two programs can be compared, even should they be written in different languages, it does have the limitation that the source code needs to be manually amended and also an acknowledged issue that the processing may require a large amount of disk space [7].

Guard, DUCT and HPCVL represent important developments in relative debugging. However, each has limitations in terms of implementation or functionality. HPCVL required users to manually instrument source code to function and DUCT is not a fully automated system. The relative debugging techniques previously described by Abramson, Sosic and Watson [3] and in the related studies are labour intensive. It is not practicable to apply such techniques to large scientific modelling software, such as the WRF model considered in this paper. These earlier techniques require deep knowledge of the target program, intuition and some luck to succeed in identifying areas where problems are likely to occur. There is a danger that the users will only find the errors they expect to find and will miss the unexpected, such as compiler errors and differences. The technique described here is able to look for differences in the entire program, even programs as large as WRF, with no need for manual intervention. It is also able to monitor execution coverage automatically so that test runs can be fine-tuned to maximise coverage of the code as it is debugged.

The functionality of Guard is centred on comparison of versions of programs. In any non-trivial program which integrates floating point values, numerical drift will cause the results from different systems to diverge. Meaningful comparisons between runs can only be made if this drift is removed. It is therefore essential that this issue is addressed. The technique described here removes numerical drift.

## 3. WEATHER RESEARCH AND FORECASTING MODEL

WRF, the Weather Research and Forecasting code developed at UCAR is arguably the most important computer program ever written [12]. It is the most widely used modelling code to inform international climate change policy, to which over 100 Billion US Dollars has been committed to date.

WRF is almost completely written in Fortran 95. The code can be built for a range of architectures, from a single serial processor to very large multi-processor systems and with varying patterns of shared memory. Parts of the code are modified by the build procedure to adapt the target architecture. The Fortran code for a serial single processor system occupies 360,410 lines. There is extensive use of Fortran 90 modules which encapsulate different areas of the computational process.

The WRF code is divided into three layers [12]. The Mediation layer describes a grid which represents an area of the Earth's atmosphere. The Model layer describes the physics of the atmosphere, and deals with issues such as mass flow, radiation, chemistry and effects at the land and sea surface. The middle layer (the Driver layer) is concerned with distributing data between the different processors and threads in the target architecture. The data structures in the top two layers are mostly large hierarchies of Fortran 95 derived types, some with their own overloaded operators. The data structures in the physics layer are mostly simple arrays of native Fortran types. The translation takes place in the middle layer. This architecture has implications for the analysis technique described in this paper.

The version of WRF used in the study was 3.1.

## 4. WinFPT Toolkit

The tool used to insert the trace statements into the source code is WinFPT. WinFPT is written and maintained by two of the authors. It analyses Fortran codes in the same way as a compiler, and stores an internal representation of the code, including comments and formatting information. A full static semantic analysis is carried out, and the tool therefore identifies the data types and other attributes of all variables, and the intents (Whether input, output both or neither) of all sub-program arguments. The trace statements are inserted into the code by modifying the internal representation. A new version of the code is generated from the internal representation. Processing of WRF by WinFPT takes about four minutes on a typical PC. The process is fully automated.

The trace file generation is handled by a small Fortran library linked with the modified code. Tools were also written to handle the large trace data files.

## 5. Removal of Numerical Drift

The technique is as follows:

a) statements are added to the code to capture every quantity produced as the result of a computation. This process is entirely automatic, and is carried out by the WinFPT tools described above. The statements inserted are subroutine calls. The results, a trace of the program flow and all intermediate data, are captured to a text file;

b) the code is run in the first computing environment, and the trace file is generated;

It is possible to run the program in the second environment, to generate a second file and to compare the two traces. This was the original intention of the authors. The results show a very large number of differences as a result of numerical drift and are almost impossible to analyse. Instead:

c) the program is run in the second environment, but the subroutines which captured the trace data in the first run are used to read the original trace file. The values computed are compared with the values read from the first run. There are four possible consequences:

1. the results are identical. No action is taken;
2. the results are similar (The criteria for similarity are passed as parameters to the program). The values computed are overwritten by the values read from file from the first trace. This prevents numerical drift;
3. the results are significantly different. The values and the point in the code where the difference occurred are recorded for investigation. The value computed is overwritten by the value from the first run so that the analysis may continue;
4. the program follows a different path through the code. The difference is recorded and the run halts. At present the situation is not recoverable.

### 5.1 Modifications to the Code

Subroutine calls are inserted automatically into the code to capture or to compare the trace data. For example, a code fragment from WRF

```
REAL, PARAMETER :: karman = 0.4
rdt = 1.0 / dtpbl
DO k = 1, kte
 sigmaf(k-1) = znw(k)
ENDDO
sigmaf(kte) = 0.0
```

is modified to:

```
REAL,PARAMETER :: karman = 0.4
CALL trace_start_sub_program('ACMPBL',6914)
rdt = 1.0/dtpbl
CALL trace_r4_data('RDT',rdt,38749)
DO k = 1,kte
 CALL trace_i4_data('K',k,38750)
 sigmaf(k-1) = znw(k)
 CALL trace_r4_data('sigmaf(k-1)',sigmaf(k-1),38751)
ENDDO
sigmaf(kte) = 0.0
CALL trace_r4_data('sigmaf(kte)',sigmaf(kte),38752)
```

There is a separate data trace subroutine for each primary data type. The arguments are a string, which records the variable assigned, including the field and array indices if any, the variable itself and a unique integer identifier which is used to locate the statement in the code. This identifier is used to check that the same path is followed through the code on each run.

### 5.2 Criteria for Similarity

Values from the two runs may differ as a result of numerical drift, and no report is made if they are sufficiently similar. The criteria for similarity are controlled as follows:

i. logical and integer values are required to match exactly;

ii. character values may be required to match, but it was found that there are relatively few character computations in the codes which have been analysed, and most of these are file names which are expected to differ,  Provision is therefore made to ignore character variables;

iii. real and complex values are required to differ by less than a percentage difference passed as a parameter to the modified program. In the study of WRF, this criterion was set to 0.1%. However, values close to zero are likely to generate large percentage differences by chance.

Therefore values were also considered to be similar if they differed by less than an absolute criterion, also passed as a parameter. In the study of WRF this was set to 1.0E-10.

### 5.3 Trace Files

The trace is written to a single ASCII file. The records, which record the data assignments, contain the identifier number, the descriptive string and the value. Values are written in a consistent format, irrespective of the Fortran kind of the data type in order that runs using different data kinds may be compared. Records are also written to mark the entry to and return from each routine.

The trace files are large. A file produced by a very short run of WRF may be of the order of 30 gigabytes. These files are too large to be handled by most text editors, and special tools were written to extract data from them. The files are also slow to produce. The present implementation of the technique requires the code to be run serially, on a single processor, and the run speed is of the order of ten times slower than the serial un-modified code.

## 6. WRF CASE STUDY

WRF may be run as a single processor serial program under Linux. Differences are observed between the output data generated by runs under GNU gfortran and Intel ifort . Some differences are due to numerical drift, but the technique also reveals coding and compiler errors. Three examples of the findings are described below.

### 6.1 Preparing WRF for Analysis

The build procedure for WRF modifies the code, and generates code automatically to adapt to different computing environments. Code is modified by the sed stream editor and by the cpp C pre- processor. This is necessary because the code must be optimised for different parallel architectures with different arrangements of shared memory. WinFPT cannot analyse the distributed code because it cannot interpret the cpp directives or anticipate the modifications made by sed. The WRF build procedure was therefore interrupted to capture the post-processed Fortran 95 files for analysis. A modified version of the build procedure was written to complete the build from the post-processed files.

### 6.2 Case 1: An Uninitialised Variable

The WRF code was processed to insert the trace statements, built with gfortran and a short run was made. The code was rebuilt with ifort and run again. The first difference reported is:

```
!*** Trace value error:
!Value computed:       148443456
!Trace file line:      24468   NEW_DOMDESC   =   538976288
```

The value computed for NEW_DOMDESC by:

ifort    is    148443456

gfortran is    538976288

The identifier, 24468, is used to identify the statement in the code where the variable was assigned. The statement is a subroutine call, and the variable should be assigned as an INTENT(INOUT) argument. However, it is never initialised and is not written to in the routine.

This is a coding error. The variable is used in the subsequent code. However it is not clear that it has any significant effect on the program results.

### 6.3 Case 2: A Coding Error, a Compiler Bug and a Poor Fortran 95 Language Feature

The comparison ifort run halts with the report:

```
!*** Trace error at start of sub-program: ESMF_TIMECOPY
!Trace file line:     9608   STOPTIME              =              0
```

ifort starts execution of the subroutine esmf_timecopy.

gfortran does not.

1)      The Coding Error

The subroutine esmf_timecopy is in esmf_time.f90 in the MODULE esmf_timemod.  It is used to overload the assignment of variables of the derived type esmf_time. The code is:

```
!===============================================================
!
!     ESMF Time Module
      module ESMF_TimeMod
!
!===============================================================
                :
! !PRIVATE TYPES:
      private
!---------------------------------------------------------------
!     ! ESMF_Time
            :
!---------------------------------------------------------------
!BOP
! !INTERFACE:
      interface assignment (=)
! !PRIVATE MEMBER FUNCTIONS:
      module procedure ESMF_TimeCopy
!  ! DESCRIPTION:
!    This interface overloads the = operator for the
!    ESMF_Time class
!
!EOP
      End interface
!
!---------------------------------------------------------------
```

There are many calls, but they are written as assignment statements. For example:

```
clockint%currtime = starttime
```

The coding error is that there is a leading PRIVATE statement in the module esmf_clockmod. esmf_timecopy is declared public, and is therefore visible when the module is used. However, ASSIGNMENT(=) is not declared public. Therefore the overloading of the operator should not be exported.

2)      The Compiler Bug

This code works as intended under ifort because of a bug in the ifort compiler. ifort exports the use of the SUBROUTINE esmf_timecopy to overload assignment when it exports the routine itself. It should not[5]. The code works as written under gfortran, not as intended.

3)      A Poor Language Feature

This is a serious failure in the Fortran 95 language. If an error occurs in the coding of an INTERFACE ASSIGNMENT (=) statement, as occurred here, then the values of the variables on which the overloaded assignment is intended to operate are simply, and silently copied.

## 6.4 Case 3: Different Orders of Computation (Not an Error)

The trace report is:

```
!*** Trace sequence error:
!Sequence number reached:         2632
!Trace line: 44475387 2444 Start sub-program:
!                                 WRF_PUT_DOM_TI_INTEGER
44475387        2632   Start sub-program: USE_PACKAGE
```

| gfortran called | `wrf_put_dom_ti_integer` |
| ifort called | `use_package` |

The code here is:

```
IF ((use_package(io_form)==io_netcdf) .OR.       &
    (use_package(io_form)==io_phdf5) .OR.        &
    (use_package(io_form)==io_pnetcdf)) THEN
   CALL wrf_put_dom_ti_integer(fid,'SURFACE_INPUT_SOURCE', &
   surface_input_source,1,ierr)
```

Both gfortran and ifort are optimising compilers. Both will evaluate each alternative of the .OR. clause and will stop evaluation as soon as one is true. The Fortran standard explicitly states that the order of evaluation is not defined by the language in this context [5]. The gfortran compiler evaluated the first sub-clause first. The ifort compiler did not, and therefore invoked the function use_package again.

This is not an error, provided that the function involved has no side effects (which is the case here). However, it is not a good idea to have an undefined pattern of program flow, and the trace analysis traps it.

The present library routines halt the program when the program flow in the second run differs from that in the first. The code was therefore modified for future runs:

```
i_use_package = use_package(io_form)
IF ((i_use_package==io_netcdf) .OR.         &
    (i_use_package==io_phdf5) .OR.          &
    (i_use_package==io_pnetcdf)) THEN
   CALL wrf_put_dom_ti_integer(fid,'SURFACE_INPUT_SOURCE', &
   surface_input_source,1,ierr)
```

### 6.5 Further Analysis of WRF

A coverage analysis of the WRF code indicated that only 18% of the code was visited in the runs which were analysed. This is to be expected. In a single WRF run, only one of 14 possible physics regimes is selected, and the atmospheric physics code is the largest part of the program. The intention of future work on the project is to analyse all of the physics code. There are also many static analyses carried out by WinFPT which expose features in the code which should be corrected.

## 7. IMPLICATIONS FOR WRF

The WRF modelling system is complex and exploits many Fortran capabilities that permit compile-time checking, such as modules and interfaces. The code is largely Fortran compliant, which is gauged by the number of successful compiler ports: IBM xlf, PGI, Intel, gfortran, g95. While the architecture of the WRF model is sound, over the years capabilities and code have been added through accretion by many different developers and contributors of the modelling community. The complicated nature and background of the source code means that the software has accumulated inconsistencies and inefficiencies with respect to the strict Fortran standards.

The initial tangible benefit that the WinFPT analyses provide is run-time stability. The FPT group is working on identifying coding errors in the WRF model. For example, they have already identified improper usages of an overloaded "=" when ESMF calls are made, and a returned value into a parameter. These errors constitute the visible portion of the iceberg for unknown, lurking bugs that could trigger failed forecasts for no apparent reason. These problems are also likely to be a source of different behaviour when running on different machines, such as when porting from one OS/compiler to another. Just as the WRF community greatly benefits from having thousands of users investigate the code and identify problems in the physical parameterization schemes, so the community will also benefit from having a tool that identifies (for later correction) errors in the Fortran code which will reduce the exposure to model failures.

With the broad usage of the WRF code across many different hardware and software systems, the intercomparability of model results becomes tantamount. Here also the WinFPT tool is able to provide a solution by allowing a fine-grained granularity comparison of results while the model is integrating.

## 8. CONCLUSIONS

The technique described isolates coding and compiler errors which lead to differences in performance when the same program is run in different computing environments. It has the advantage that it is highly automated, and requires very little human intervention. However, it is computationally intensive, and currently requires code to be run serially on a single processor.

Initial investigation making use of the techniques described has enabled a number of issues to be discovered. These issues can be related to coding in the model and the compilers used to build the code. The paper discusses three findings from the experiments that have been performed. These represent important validation of this technique, and the intention of the project is to apply the technique to include broader coverage of the WRF model code.

The initial experiments have focussed on single processor builds of the WRF model. By making use of alternative test suites, it is envisaged that coverage of the WRF code will be significantly increased from the 18% which has currently been achieved. However, this could be further improved by expanding the experimentation to also include multiprocessor builds. One aim of future work on the project is to investigate the combination of aspects of this technique and the relative debugging techniques developed by Abramson et al. [2] to analyse runs made on multiprocessor systems.

The technique described here offers the possibility of a new phase in software development in which software validation can be enhanced by comparing execution on different compilers to track down bugs that would be difficult or impossible to detect otherwise. This is especially useful where immature compilers are involved.